\documentclass[aps,prl,twocolumn,nofootinbib,superscriptaddress]{revtex4-2}

\usepackage{amsmath,amssymb,bm,graphicx}
\usepackage[colorlinks=true,citecolor=blue,linkcolor=blue]{hyperref}

\begin{document}

\title{Experimental Access to the Gluonic Origin of the Proton Mass}

\author{Zein-Eddine Meziani}
\affiliation{Physics Division, Argonne National Laboratory, Lemont, Illinois, 60439, USA}
\author{Ismail Zahed}
\affiliation{Center for Nuclear Theory, Department of Physics and Astronomy,
Stony Brook University, Stony Brook, New York, 11794, USA}

\date{\today}

\begin{abstract}
Most of the proton mass originates not from the Higgs mechanism but from the quantum structure of the QCD vacuum.
The dominant contribution arises from the gluonic trace anomaly associated with the breaking of conformal symmetry in quantum chromodynamics.
We show that this anomaly contribution is experimentally accessible through scalar gravitational form factors.
The key observable is the scalar gluonic trace form factor of the proton, which can be reconstructed from three measurable quantities: the quark scalar gravitational form factor accessible in deeply virtual Compton scattering, the gluon scalar gravitational form factor measurable in near-threshold heavy quarkonium production, and the nucleon sigma-term form factor.
We also show that the scalar gluonic form factor extracted from the trace anomaly is quantitatively consistent with lattice QCD and the instanton liquid model over hadronic distance scales.
These results provide a direct experimental path to probing the gluonic origin of visible mass.
\end{abstract}

\maketitle


{\it Introduction.} -
Understanding the origin of the proton mass is a central problem in quantum chromodynamics (QCD).
In the limit of massless light quarks, the QCD Lagrangian contains no intrinsic mass scale, yet the proton remains massive.
The small current quark masses generated through the Higgs mechanism account for only a minor fraction of the proton mass.
Most visible mass therefore emerges dynamically from the quantum structure of the QCD vacuum.

The underlying mechanism is the quantum breaking of scale invariance through the trace anomaly of the QCD energy-momentum tensor (EMT)~\cite{Crewther:1972kn,Collins:1977jy,
Vainshtein:1981wh},
\begin{equation}
T^\mu_{\ \mu}
=
\frac{\beta(g)}{2g}F^a_{\mu\nu}F^{a\mu\nu}
+
\sum_f(1+\gamma_m)m_f\bar\psi_f\psi_f .
\label{trace}
\end{equation}
Here
$F^{a\mu\nu}$ is the gluon field-strength tensor and~\cite{Gross:1973id,Politzer:1973fx}
\begin{equation}
\beta(g)=\mu\frac{dg}{d\mu}
\end{equation}
is the QCD beta function governing the running of the strong coupling $g$.
The second term in Eq.~(\ref{trace}) is the explicit light quark-mass contribution with the quark anomalous dimension \(\gamma_m\simeq 0.295\) (at 2 GeV).

A major challenge has been to identify experimentally measurable observables that isolate the anomaly contribution.
Recent progress in gravitational form factors, deeply virtual exclusive reactions,
and near-threshold heavy quarkonium production has opened a realistic path toward such measurements~\cite{Burkert:2018bqq,Kumericki:2019ddg,Duran:2023het,Adhikari:2023yft,CLAS:2026lls,007:2026dow}. The importance of these measurements for the case of deeply virtual Compton scattering (DVCS) were advocated theoretically by many~\cite{Ji:1996nm,Radyushkin:1996nd,Diehl:2003ny,Belitsky:2005qn}. In the case of near-threshold $J/\psi$ production on the nucleon, recent  measurements were initially motivated in part by the discovery of the LHCb charm pentaquark~\cite{LHCb:2015yax,Karliner:2015voa,Kubarovsky:2015aaa,Brodsky:2000zc, Hafidi:2017bsg} However, the use of such measurements to probe the gluonic structure of the proton had already been proposed earlier~\cite{Kharzeev:1995ij,Kharzeev:1998bz,Brodsky:2000zc,Hafidi:2017bsg}.

In this work we show that the scalar gluonic trace form factor of the proton,
which governs the dominant contribution to the proton mass,
can be reconstructed experimentally from three measurable quantities:

\begin{itemize}
\item[(i)]
The quark scalar gravitational form factor accessible in deeply virtual Compton scattering (DVCS)~\cite{Burkert:2018bqq,Kumericki:2019ddg};

\item[(ii)]
The gluon scalar gravitational form factor measurable in near-threshold heavy quarkonium production~\cite{Duran:2023het,Adhikari:2023yft,CLAS:2026lls};

\item[(iii)]
The scalar sigma-term form factor associated with explicit quark masses~\cite{Steele:1997ms,Hoferichter:2012wf}.
\end{itemize}

We further show that the resulting scalar gluonic form factor is quantitatively consistent with lattice QCD and with the instanton liquid model, indicating that the scalar gluonic structure of the proton is dominated by semiclassical topological glue over hadronic distance scales.

We use the symmetric gauge-invariant decomposition~\cite{Ji:1995sv,Ji:1995pc,Ji:1997pf,Ji:2025xxx,Tanaka:2018nae,Polyakov:2018zvc}
\begin{equation}
T^{\mu\nu}=T_q^{\mu\nu}+T_g^{\mu\nu},
\end{equation}
where the full EMT is conserved,
\begin{equation}
\partial_\mu T^{\mu\nu}=0,
\end{equation}
while the separate quark and gluon pieces satisfy
\begin{equation}
\partial_\mu T_q^{\mu\nu}
=
-\partial_\mu T_g^{\mu\nu}.
\end{equation}
Their trace is constrained by the anomalous Ward identity,
Eq.~(\ref{trace}).
The explicit operator definitions of the EMT are summarized in Appendix A.


{\it Scalar gravitational form factors and the trace anomaly.} -
For a spin-averaged nucleon state,
the nonconserved scalar component of the quark and gluon EMTs can be parameterized as~\cite{Tanaka:2018nae,Polyakov:2018zvc}
\begin{equation}
\langle p'|T_a^{\mu\nu}|p\rangle_{\rm nc}
=
2M^2g^{\mu\nu}\bar C_a(t),
\qquad a=q,g,
\end{equation}
where
$t=\Delta^2$
with
$\Delta=p'-p$,
and $M$ is the proton mass.
Energy-momentum conservation requires
\begin{equation}
\bar C_q(t)=-\bar C_g(t).
\end{equation}

The scalar gravitational form factor is defined 
through~\cite{Ji:2021znw,Mamo:2021krl,Mamo:2022eui,Ji:2025xxx}
\begin{equation}
G_s(t)=G_{s,q}(t)+G_{s,g}(t)
\end{equation}
\begin{equation}
G_{s,a}(t)
=
A_a(t)
+
\frac{t}{4M^2}B_a(t)
-
\frac{3t}{M^2}C_a(t),
\end{equation}
where
$A_a(t)$,
$B_a(t)$,
and
$C_a(t)$
are the standard gravitational form factors entering the nucleon EMT decomposition~\cite{Ji:1995sv,Ji:1995pc,Polyakov:2018zvc}, and $M$ the nucleon mass.

Taking the trace of the quark EMT gives
\begin{equation}
\langle p'|T^\mu_{q\mu}|p\rangle
=
M\bar U(p')U(p)
\left[
G_{s,q}(t)+4\bar C_q(t)
\right].
\label{qtrace}
\end{equation}
The nucleon spinors are normalized as \(\bar U(p')U(p)=2M.\)
Using Eq.~(\ref{trace}),
this matrix element equals the scalar quark operator,
\begin{equation}
\langle p'|\sum_f(1+\gamma_m)m_f\bar\psi_f\psi_f|p\rangle
=
\sigma(t)\bar U(p')U(p),
\label{sigmaff}
\end{equation}
where
$\sigma(t)$
is the generalized sigma-term form factor, hence
\begin{equation}
4\bar C_g(t)
=
G_{s,q}(t)-\frac{\sigma(t)}{M}.
\label{CgGsSigma}
\end{equation}

The gluonic contribution to the trace anomaly satisfies
\begin{equation}
\frac{
\langle p'|T^\mu_{\ g\mu}|p\rangle
}{
M\bar U(p')U(p)
}
=
G_{s,g}(t)+4\bar C_g(t).
\end{equation}
Using Eq.~(\ref{CgGsSigma}) gives the central result
\begin{equation}
\frac{
\langle p'|T^\mu_{\ g\mu}|p\rangle
}{
M\bar U(p')U(p)
}
=
G_{s,g}(t)
+
G_{s,q}(t)
-
\frac{\sigma(t)}{M}.
\label{main}
\end{equation}
Equation~(\ref{main}) shows that the scalar gluonic trace form factor,
which governs the dominant contribution to the proton mass,
can be reconstructed experimentally from measurable scalar gravitational form factors together with the sigma-term form factor.

\begin{figure}[b!]
\centering
\includegraphics[width=\columnwidth]{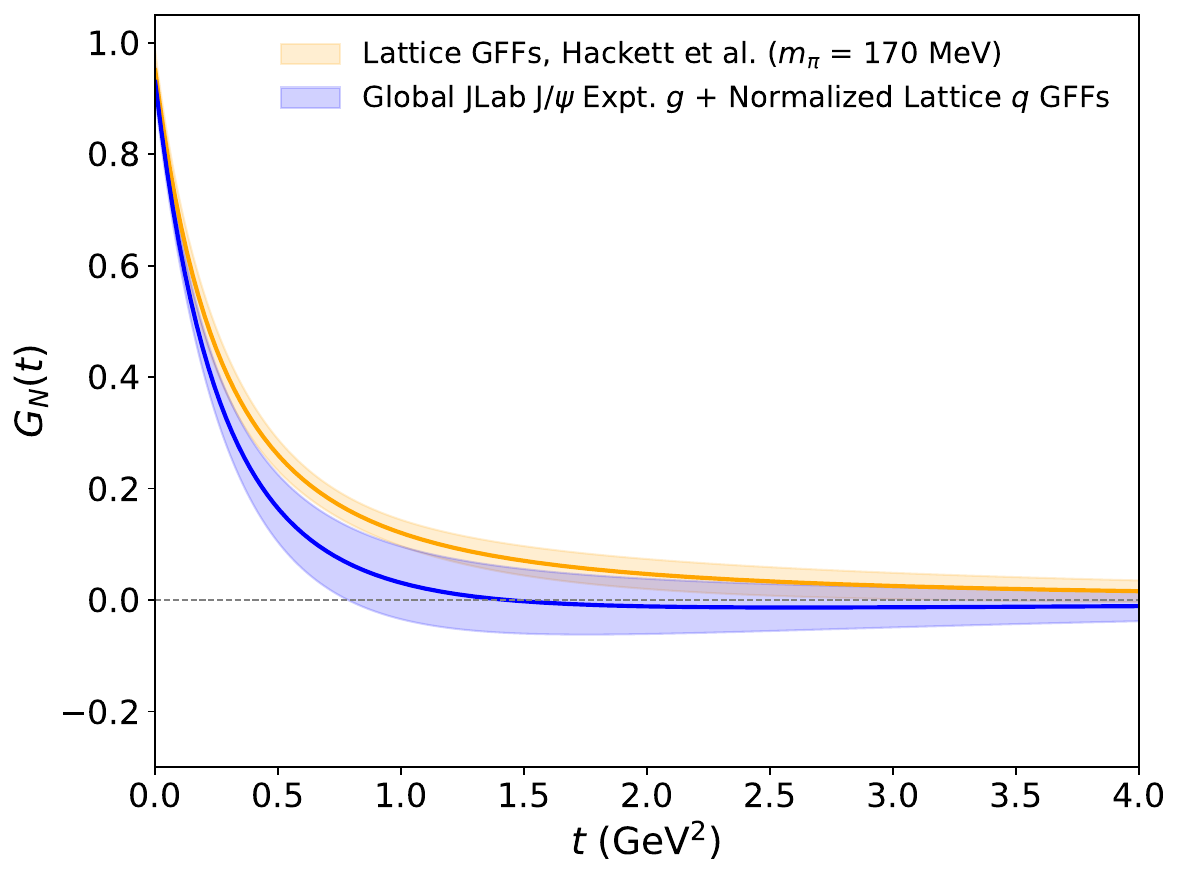}
\caption{
Scalar gluonic form factor of the proton,
$G_N(t)$,
obtained from the experimental reconstruction of
Eq.~(\ref{GNmeas}).
The gluon contribution is extracted from the global analysis of
the near-threshold $J/\psi$ production from the proton experiments~\cite{Duran:2023het,Adhikari:2023yft,CLAS:2026lls,007:2026dow} at Jefferson Lab using holographic QCD~\cite{Mamo:2022eui}.
The quark contribution is a dipole/tripole fit to $A_q(t)$/$C_q(t)$ lattice form factors, with the corresponding quark EMT normalized such that $A_q(0)$ matches the phenomenological $A_q(0)=0.576\pm 0.011$~\cite{Hackett:2024gff,pefkou_private,Hou:2019efy}
The resulting scalar gluonic form factor (blue band) is compared
with the independent lattice-QCD determination obtained from the
gravitational form factors~\cite{Hackett:2024gff}
(orange band).
The experimentally reconstructed
trace-anomaly form factor is in fair agreement with the scalar
structure obtained directly in lattice QCD.
}
\label{fig:GFFcomparison}
\end{figure}

{\it Scalar gluonic form factor.} -
The scalar gluonic form factor of the proton is defined through the matrix element of the gluonic trace operator,
\begin{equation}
\frac{\beta(g)}{2g}
\langle P'|F^2|P\rangle
=
M_NG_N(Q^2)\bar u(P')u(P),
\label{GNdef}
\end{equation}
where
$Q^2=-t$
and
$F^2\equiv F^a_{\mu\nu}F^{a\mu\nu}$.
Physically,
$G_N(Q^2)$
measures the scalar gluonic structure of the proton associated with the trace anomaly.
At zero momentum transfer,
\begin{equation}
G_N(0)=1-\frac{\sigma(0)}{M_N},
\end{equation}
Equation~(\ref{main}) provides direct experimental access to this quantity through the measurable scalar EMT combination
\begin{equation}
G_N(t)
=
G_{s,g}(t)
+
G_{s,q}(t)
-
\frac{\sigma(t)}{M}.
\label{GNmeas}
\end{equation}

In the instanton vacuum,
the scalar gluonic operator measures fluctuations in the density of topological tunneling gauge fields permeating the QCD vacuum,
\begin{equation}
\frac{g^2F^2}{32\pi^2}
\rightarrow
\frac{N_++N_-}{V},
\label{eq:F2F2}
\end{equation}
where
$N_+$
and
$N_-$
denote the numbers of instantons and antiinstantons in the space-time volume $V$~\cite{Schafer:1996wv,Diakonov:2002fq,Nowak1996,Shuryak:2026pqt}.
A nucleon propagating through this medium acts as a localized color source that polarizes and partially depletes the surrounding gluonic vacuum fluctuations~\cite{Zahed:2021fxk}.
The scalar gluonic form factor therefore probes how the proton reshapes the underlying topological vacuum structure responsible for the emergence of hadronic mass.

This interpretation is to be contrasted with recent continuum bound states calculations of the proton gravitational form factors~\cite{Yao:2024ixu}, where the scalar response is obtained from the dressed quark core and its gluonic interactions. Here the same observable is instead viewed as a probe of the polarization of the topological QCD vacuum surrounding the nucleon. For completeness, we also note the dispersive analyses of these form factors in~\cite{Cao:2024zlf,Cao:2025dkv}.


\begin{figure}[t]
\centering
\includegraphics[width=\columnwidth]{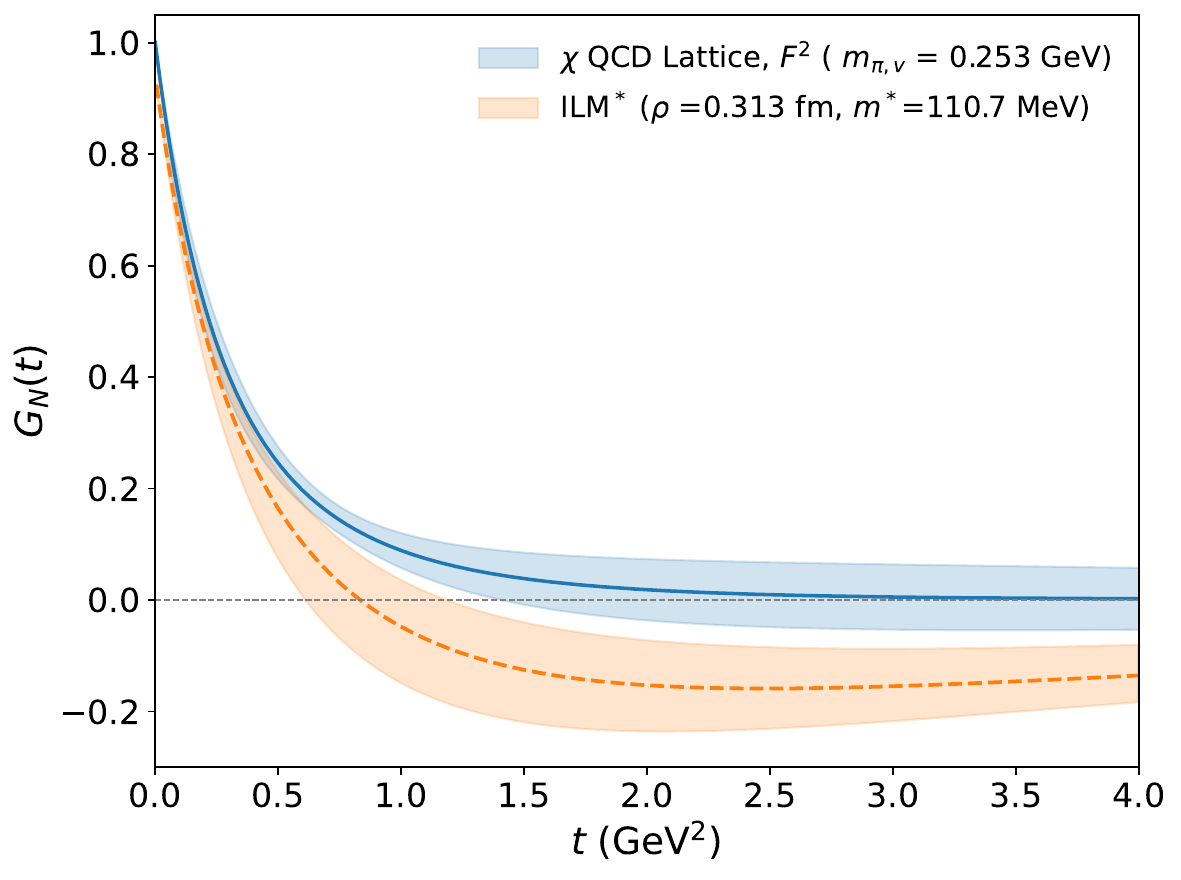}
\caption{
Comparison of the scalar gluonic form factor obtained from
the trace-anomaly lattice calculation
(blue band)~\cite{Wang:2024lrm}
with the prediction of the interacting instanton liquid model
(orange band)~\cite{Liu:2024rdm}.
}
\label{fig:ILMcomparison}
\end{figure}

The comparison in Fig.~\ref{fig:GFFcomparison} shows that the
experimentally reconstructed scalar gluonic form factor falls rapidly
with momentum transfer. In the low-momentum region
\(-t\lesssim 1\,{\rm GeV}^2\), this behavior is close to the
instanton-liquid result shown in Fig.~\ref{fig:ILMcomparison}.
This is precisely the domain in which the instanton description is
expected to be meaningful, since it probes the long-distance
topological component of the QCD vacuum rather than short-distance
perturbative gluon dynamics.

In Fig.~\ref{fig:ILMcomparison} and below \(-t\simeq 1\,{\rm GeV}^2\), the ILM form factor
has the same rapid fall-off as the experimentally extracted
\(G_N(t)\) in Fig.~\ref{fig:GFFcomparison}. At larger momentum
transfer, the ILM curve departs from the lattice trace-anomaly result,
as expected once shorter-distance gluonic correlations beyond the
semiclassical instanton medium become important.

{\it Conclusions.} --

We have shown that the dominant gluonic contribution to the proton mass, encoded in the QCD trace anomaly, can be determined experimentally through scalar gravitational form factors.

Our chief  result, Eq.~(\ref{main}), expresses the scalar gluonic trace form factor entirely in terms of three measurable quantities: the quark scalar gravitational form factor accessible in deeply virtual Compton scattering, the gluon scalar gravitational form factor measured in near-threshold heavy-quarkonium production, and the nucleon sigma-term form factor. This relation provides a direct experimental determination of the scalar gluonic form factor associated with the trace anomaly and therefore with the dominant origin of the proton mass.

The reconstructed scalar gluonic form factor has been compared with independent lattice-QCD calculations and with the interacting instanton liquid model. The comparison with lattice calculations provides a nontrivial consistency check of the experimental reconstruction, while the agreement with the instanton prediction over the low-momentum region,
\(-t\lesssim1~{\rm GeV}^2\), where the instanton liquid model is expected to be applicable, indicates that the long-distance scalar gluonic structure of the proton is dominated by semiclassical topological gauge fields. At larger momentum transfers, additional short-distance gluonic dynamics are expected to become important.

These results establish a direct connection between precision measurements of scalar gravitational form factors and the nonperturbative gluonic structure of the proton. They provide an experimentally accessible framework for quantifying the trace anomaly, probing the gluonic origin of visible mass, and investigating the role of topological vacuum fluctuations in quantum chromodynamics.


{\it Acknowledgements.} --
We thank Xiangdong Ji for discussions. ZEM thanks Dimitra Pefkou for providing the lattice dipole-tripole fit parameters for A-D GFFs, Bigeng Wang and Keh-Fei Liu for providing the results of the trace anomaly lattice calculations.

This work is supported by the U.S. Department of Energy,
Office of Science, Office of Nuclear Physics under Contract No.~DE-FG-88ER40388 (IZ) and DE-AC02-06CH11357 (ZEM). 
This work is also supported in part by the Quark-Gluon Tomography (QGT)
Topical Collaboration under Award No.~DE-SC0023646 (IZ).


\appendix


\appendix

\section{Energy-Momentum Tensor Parametrization}
\label{AppA}
The symmetric gauge-invariant quark and gluon energy-momentum tensors are defined as
\begin{equation}
T_q^{\mu\nu}
=
\frac14
\bar\psi
\left[
\gamma^\mu i\overleftrightarrow D^\nu
+
\gamma^\nu i\overleftrightarrow D^\mu
\right]\psi ,
\end{equation}
where
$\psi$
is the quark field,
$D^\mu$
the gauge-covariant derivative,
and
$\overleftrightarrow D^\mu=\overrightarrow D^\mu-\overleftarrow D^\mu$.

The gluon contribution is
\begin{equation}
T_g^{\mu\nu}
=
-F^{a\mu\alpha}F^{a\nu}_{\ \ \alpha}
+
\frac14g^{\mu\nu}F^a_{\alpha\beta}F^{a\alpha\beta}.
\end{equation}

The nucleon matrix element of the quark and gluon EMTs admits the standard decomposition~\cite{Ji:1995sv,Ji:1995pc,Polyakov:2018zvc,Diehl:2003ny,Belitsky:2005qn}
\begin{align}
\langle p'|T_a^{\mu\nu}|p\rangle
&=
\bar U(p')
\Bigg[
A_a(t)\gamma^{(\mu}P^{\nu)}
\nonumber\\
&\qquad
+
B_a(t)\frac{P^{(\mu}i\sigma^{\nu)\alpha}\Delta_\alpha}{2M}
\nonumber\\
&\qquad
+
\frac{1}{4M}
C_a(t)
\left(
\Delta^\mu\Delta^\nu
-g^{\mu\nu}\Delta^2
\right)
\nonumber\\
&\qquad
+
M\bar C_a(t)g^{\mu\nu}
\Bigg]
U(p),
\label{fullEMT}
\end{align}
where
$a=q,g$ labels the quark and gluon contributions,
\begin{equation}
P^\mu=\frac{p^\mu+p'^\mu}{2},
\qquad
\Delta^\mu=p'^\mu-p^\mu,
\qquad
t=\Delta^2 ,
\end{equation}
and
$\gamma^{(\mu}P^{\nu)}=
(\gamma^\mu P^\nu+\gamma^\nu P^\mu)/2$.



\section{Vacuum Compressibility}

The connected scalar gluonic correlator~\cite{Nowak1996,Zahed:2021fxk}
\begin{equation}
\sigma_{F^2}
=
\frac{1}{32\pi^2}
\int d^4x\,
\frac{
\langle F^2(x)F^2(0)\rangle_C
}{
\langle F^2(0)\rangle
},
\end{equation}
defines the compressibility of the topological QCD vacuum. In the instanton vacuum,
the scalar gluonic normalization contains the fluctuation strength of the topological medium~\cite{Zahed:2021fxk}
\begin{equation}
G_N(0)
=
-\frac{4\pi^2\beta(g)}{g^3}
\frac{\sigma_{F^2}}{\bar N}
\bigg(1-\frac{\sigma(0)}{M_N}\bigg)
\end{equation}
It follows that
\begin{equation}
\frac{\sigma_{F^2}}{\bar N}
=
-\frac{g^3}{4\pi^2\beta(g)}
\approx
\frac{4}{11N_c/3-2N_f/3}
\end{equation}
using the one-loop relation.

\section{Pion-nucleon scalar FF}
The pion-nucleon sigma term form factor in the ILM can be obtained in the mean-field approximation, by resumming the t-channel bubble diagrams between the constitutive quarks induced by the emergent ’t Hooft interactions. The result in the zero size approximation is~\cite{Liu:2024jno}
\begin{equation}
\sigma_{\pi N}(t) =\frac{\langle p_2|\sum_fm_f\overline\psi_f\psi_f|p_1\rangle}{2M^2}=
\frac{\sigma_{\pi N}(0)}{1-t/m_\sigma^2}
\end{equation}
where $m_\sigma = 683$ MeV   and $\sigma_{\pi N}(0)\simeq 0.053$


\bibliographystyle{apsrev4-2}
\bibliography{G-ZM-refs}

\end{document}